\documentclass[fleqn,12pt,twoside]{article}
\usepackage{espcrc1}
\usepackage{floatflt}
\usepackage{wrapfig}

\def\PRL{Phys. Rev. Lett. }
\def\PRC{Phys. Rev. C }
\def\PRD{Phys. Rev. D }

\def\etal{\emph{et al.}}

\def\pT{\mbox{$p_T$}}
\def\v2{\mbox{$v_2$}}
\def\sqrtsNN{\mbox{$\sqrt{s_{NN}}$}}

\usepackage{graphicx}
\usepackage[figuresright]{rotating}

\newcommand{\AmS}{{\protect\the\textfont2
  A\kern-.1667em\lower.5ex\hbox{M}\kern-.125emS}}

\title{Azimuthal Anisotropy of Charged and Identified High $p_T$ 
Hadrons in Au+Au
Collisions at RHIC}

\author{K. Filimonov\address{Lawrence Berkeley National Laboratory,\\
	1 Cyclotron Road, Berkeley, California 94720, United States} 
	for the STAR collaboration\thanks{For the full author list and acknowledgements, see Appendix
"Collaborations" of this volume}  and the STAR-RICH collaboration$^*$}
       
\begin{document}

\maketitle

\begin{abstract}
We report new results on
$v_2(p_T)$ for Au+Au collisions at $\sqrt{s_{NN}}$=200 GeV 
for charged hadrons, pions, kaons, (anti)protons, $K^0_s$, 
and $\Lambda$. The analysis is extended to $p_T=12$ GeV/c for charged
hadrons and $p_T=4$ GeV/c for identified particles. A 
comparison of the azimuthal anisotropy of charged hadrons
measured at \sqrtsNN=130 and 200 GeV is presented.
The $p_T$-dependence of baryon versus meson elliptic flow 
is discussed. 

\end{abstract}

\section{INTRODUCTION}

QCD calculations predict that high energy 
partons traversing nuclear matter lose energy through induced gluon radiation 
\cite{energyloss,energyloss2}.
Recent measurements of inclusive charged hadron
distributions in Au+Au collisions at \sqrtsNN=130 GeV
show a suppression of hadron yields at high \pT\  
in central collisions relative to
peripheral collisions and scaled nucleon-nucleon interactions \cite{phenix,StarHighpt,klay},
consistent with the picture of partonic energy loss in a dense system.
 The
fragmentation products of high energy partons that have propagated 
through the azimuthally 
asymmetric
system generated by non-central collisions may exhibit azimuthal
anisotropy due to energy loss and the azimuthal dependence of the path
length \cite{wangglv}. 
Elliptic flow \cite{olli} at $p_T<2$ GeV/c 
measured at \sqrtsNN=130 GeV  reaches the 
values predicted by hydrodynamical models based on the assumption of 
complete local thermalization \cite{v2charged,v2identified}.
At high $p_T$, azimuthal distributions of charged hadrons 
exhibit a structure suggestive
of fragmentation of hard scattered partons as well as an anisotropy with 
respect to the reaction plane \cite{v2highpt}. 
The new STAR data on the azimuthal anisotropy parameter $v_2$ 
for Au+Au 
collisions at $\sqrt{s_{NN}}$=200 GeV extend the measurements 
up to $p_T=12$ GeV/c and
 provide important constraints 
on the underlying mechanisms of high $p_T$ hadron production in nuclear
collisions.

\section{EXPERIMENT AND ANALYSIS}

The main tracking detector in STAR is the Time Projection Chamber (TPC) 
situated in a solenoidal magnetic field of 0.5 T. The TPC has wide
pseudorapidity $|\eta|<1.3$ and complete azimuthal coverage, with 
excellent momentum resolution. Charged hadrons are 
identified by measuring 
specific energy loss $dE/dx$ for $p_T<1$ GeV/c. $K^0_s$ and $\Lambda$ are identified via their $V^0$-decay 
topology.
A Ring Imaging Cherenkov Detector (STAR-RICH) \cite{rich}, 
covering $\Delta\phi=20^\circ$ and $|\eta|<0.3$ and situated on the
exterior of the TPC, extends charged
hadron identification to high $p_T$.

2.7~M minimum-bias triggered Au+Au interactions 
at \sqrtsNN=200 GeV were used for this analysis.
The minimum-bias data  
correspond to $94\pm 3$\% of the geometric cross section.
All tracks used in this analysis passed within 1 cm of primary vertex and
had at least 20 measured space points. 

The azimuthal anisotropy of an event in momentum space 
is quantified by the coefficients of the  Fourier decomposition of the 
azimuthal particle distributions, with 
the second harmonic coefficient $v_2$ 
referred to as elliptic flow~\cite{voloshin}.
$v_2$ is inferred from the azimuthal particle 
distribution with respect 
to the estimated reaction plane orientation, corrected for the reaction plane 
resolution. The track selection and calculation
of the reaction plane are described in 
\cite{v2highpt}.
The reaction plane analysis integrates all possible sources of azimuthal 
correlations, including those unrelated to the orientation of the 
reaction plane. 
A four-particle cumulant method \cite{4part} for 
flow measurements reduces non-flow sources (resonance decays, 
(mini)jets, final state interactions, 
momentum conservation, etc.)
to a negligible level \cite{flowprc}. 

\section{AZIMUTHAL ANISOTROPY OF CHARGED HADRONS}

\begin{wrapfigure}[15]{l}{85mm}
\vspace{-10mm}
\includegraphics[width=85mm]{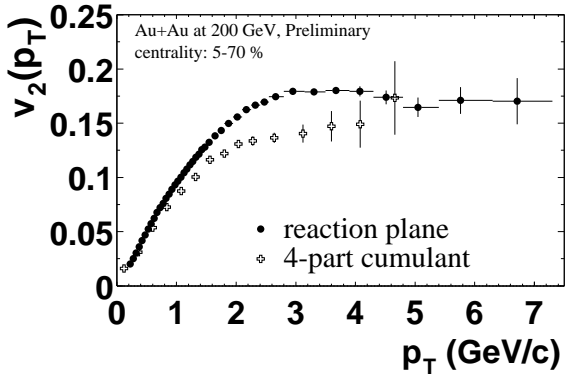}
\vspace{-7mm}
\caption{$v_2(p_T)$ of charged hadrons from 
the reaction plane (circles) 
and four-particle cumulant method (crosses) for centralities 5-70\%.}
\label{fig:4part}
\vspace{2.0mm}
\end{wrapfigure}

Figure~\ref{fig:4part} shows the $p_T$-dependence of $v_2$ from
the reaction plane method and 
four-particle cumulant method. The new results from the four-particle
cumulant method have much higher statistical precision than 
previously \cite{flowprc} 
for centralities 5-70\%. Similar to the results at 130 GeV \cite{v2highpt,flowprc}, $v_2$ rises
linearly for $p_T<1$ GeV/c, but saturates
for $p_T>3$ GeV/c for both methods.
 The $v_2$ values obtained from the 
four-particle cumulant method are 20\% lower than those from the reaction 
plane analysis.
This difference is attributed to the 
non-flow correlations in the reaction plane analysis, and 
we assign a one-sided 20\% systematic uncertainty for the reaction plane $v_2$ values presented below.

Figure~\ref{fig:ratio}, upper panel, shows $v_2(p_T)$ of charged hadrons 
obtained with the reaction plane analysis 
for minimum-bias collisions at 130 and 200 GeV.  The 
functional form of $v_2(p_T)$-de\-pendence is similar at the two energies. 
Figure~\ref{fig:ratio}, lower panel, shows the ratio of $v_2$  
from the 200 GeV and
130 GeV datasets. The elliptic flow is stronger at 200 GeV for $p_T<1$ GeV/c, suggesting a higher 
degree of thermalization.
\begin{figure}[htb]
\begin{minipage}[t]{78mm}
\includegraphics[width=78mm]{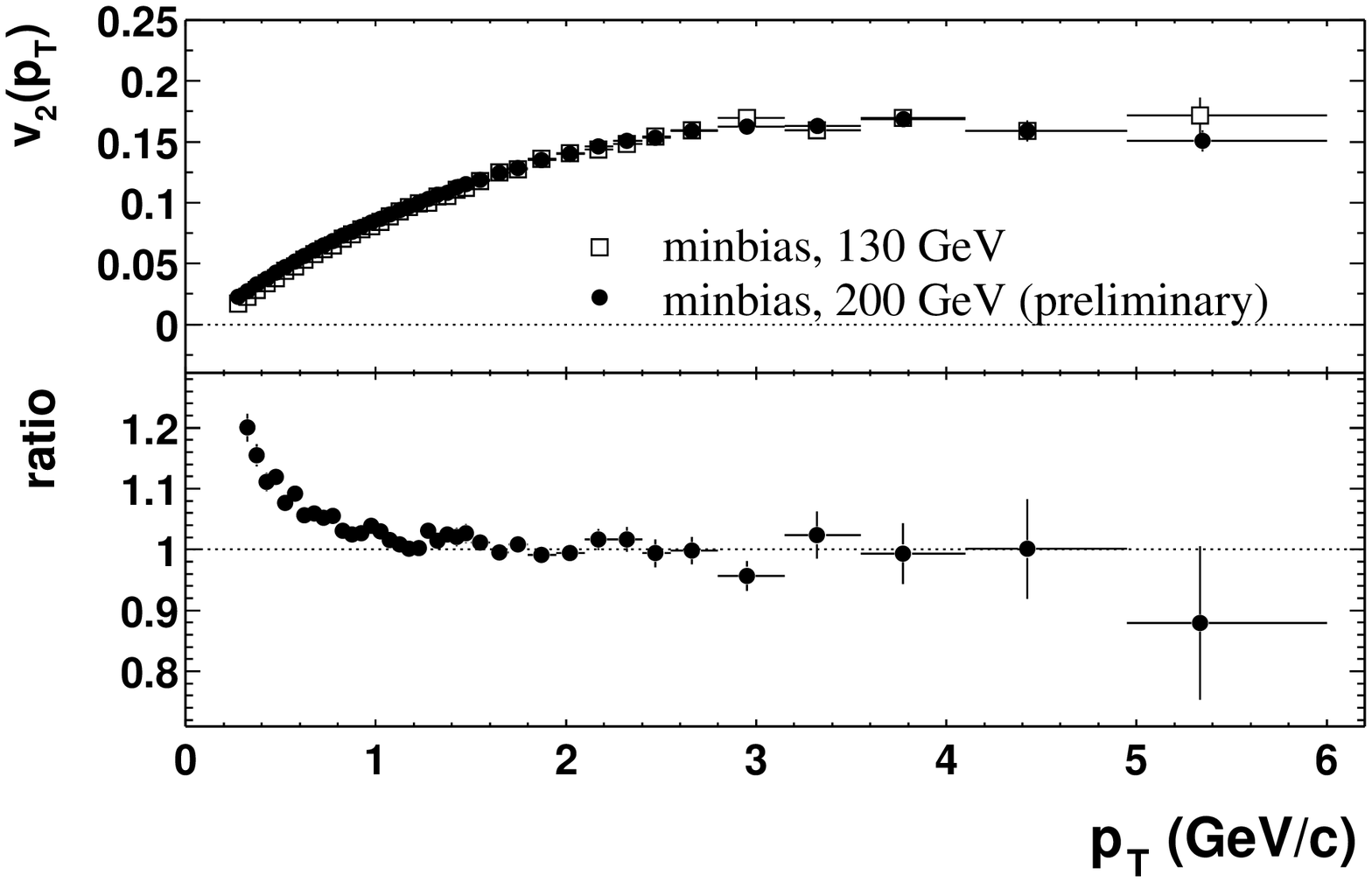}
\vspace{-17mm}
\caption{Upper panel: $v_2(p_T)$ of charged particles for minimum-bias collisions at \sqrtsNN=130 \cite{v2highpt} and 200 GeV. Lower panel: ratio of $v_2$ at 200 GeV to that at 130 GeV.}
\label{fig:ratio} 
\end{minipage}
\hspace{\fill}
\begin{minipage}[t]{78mm}
\includegraphics[width=78mm]{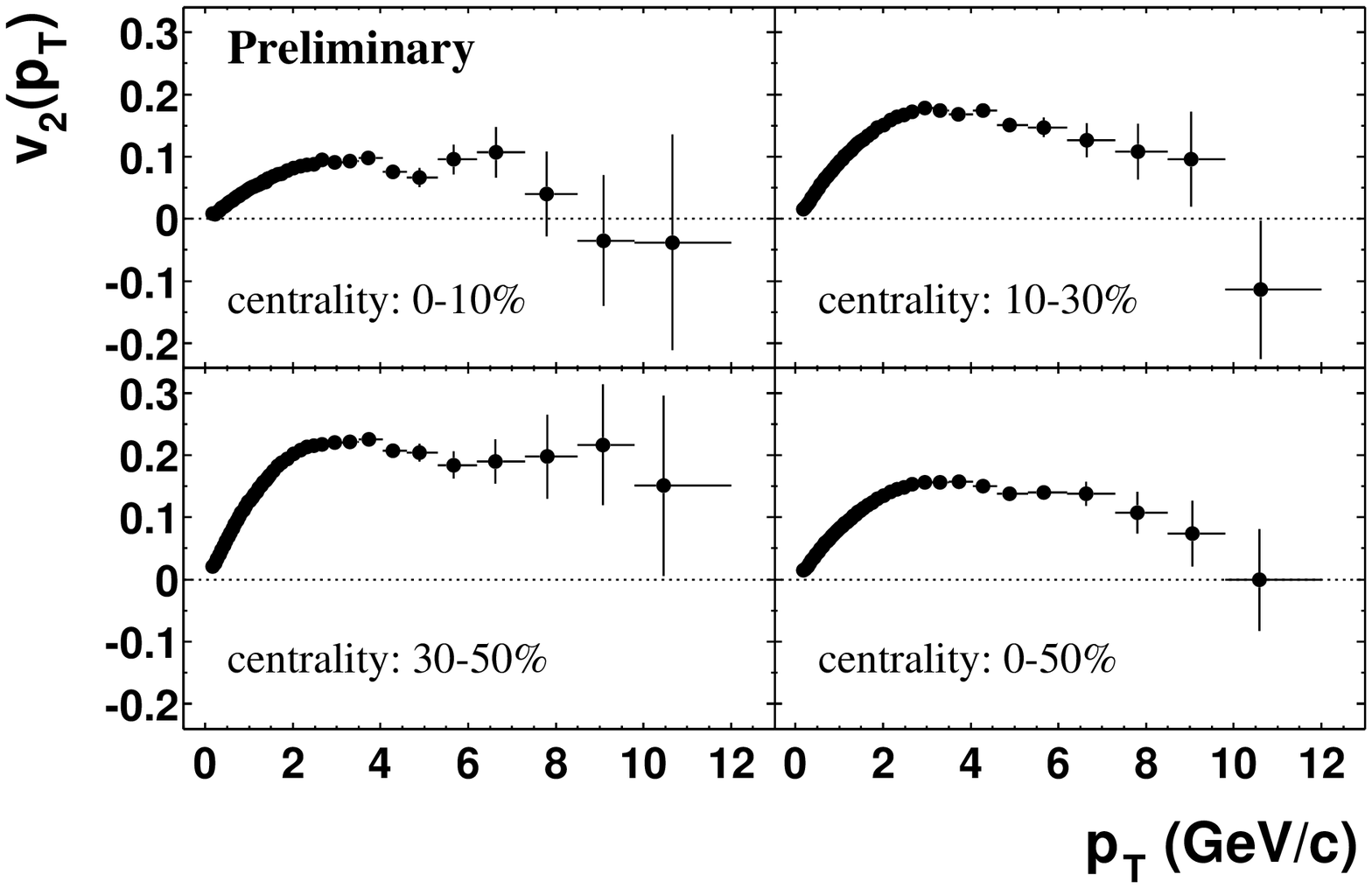}
\vspace{-17mm}
\caption{$v_2(p_T)$ of charged particles for different centralities at 200 GeV.}
\label{fig:charged}
\end{minipage}
\vspace{-5mm}
\end{figure}
In the saturation region at high $p_T$, 
the magnitude of the azimuthal anisotropy is 
the same to within 5\%, in contrast to the growth of the inclusive cross section with increasing \sqrtsNN~\cite{klay}. This is an indication of the geometrical origin 
of the large anisotropies observed in the hard scattering region.

Figure~\ref{fig:charged} shows the $v_2(p_T)$-dependence for different
collision centralities for $p_T<12$ GeV/c. 
$v_2$ remains finite for non-central collisions, exhibiting a decrease from 
the saturation level
at the highest measured $p_T$ for the more central events. It is expected
that the azimuthal anisotropy will vanish in the limit of very high $p_T$.
$v_2$ values 
measured at $3<p_T<6$ GeV/c for a restricted centrality range, corrected
for contribution from non-flow effects, 
are consistent with the maximum expected $v_2$ from surface 
emission \cite{shuryak}.

\section{ELLIPTIC FLOW OF MESONS AND BARYONS}
\begin{figure}[htb]
\vspace{-3mm}
\begin{minipage}[t]{78mm}
\includegraphics[width=85mm]{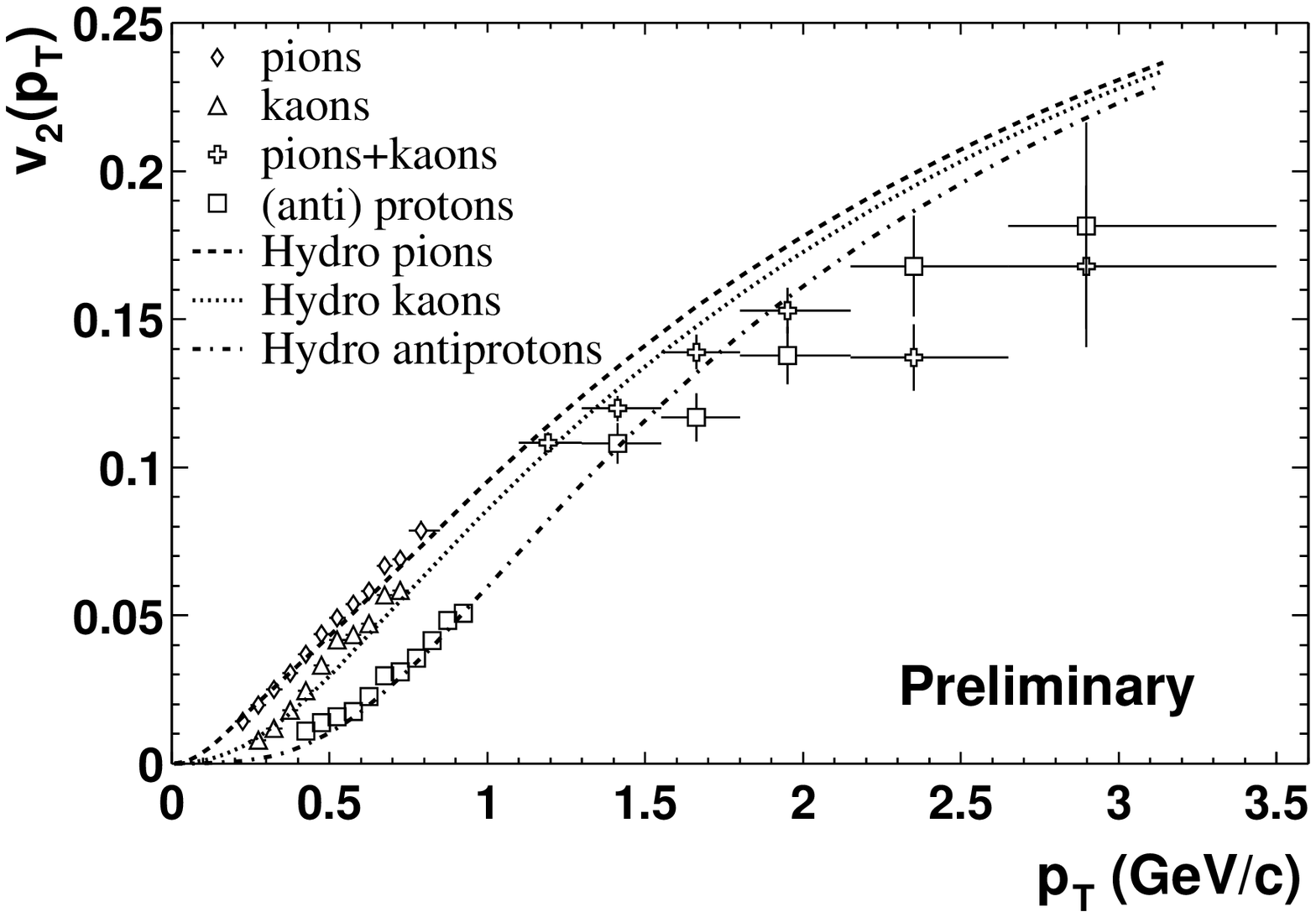}
\vspace{-13mm}
\caption{$v_2(p_T)$ of identified charged particles for minimum-bias events
at \sqrtsNN=200 GeV, compared to hydro calculations.}
\label{fig:rich}
\end{minipage}
\hspace{\fill}
\begin{minipage}[t]{78mm}
\includegraphics[width=85mm]{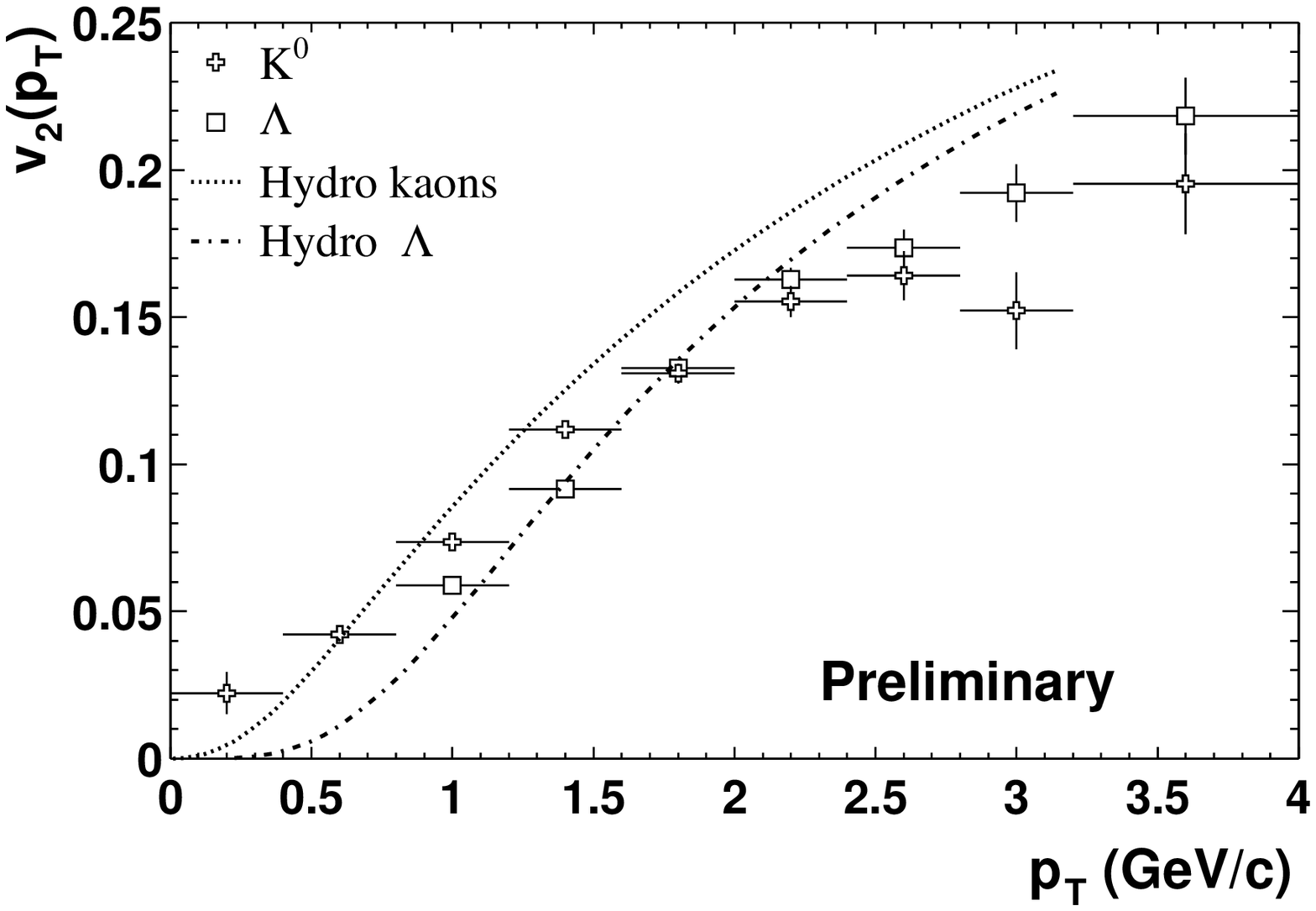}
\vspace{-13mm}
\caption{$v_2(p_T)$ of $K^0_s$ and $\Lambda$ for minimum-bias events
at \sqrtsNN=200 GeV, compared to hydro calculations.}
\label{fig:lambda}
\end{minipage}
\vspace{-0.4cm}
\end{figure}

A proposed explanation of
the saturation
of $v_2$ at transverse momenta $2<p_T<5$ GeV/c is the dominance
of baryons in this $p_T$ region described by non-perturbative
mechanisms such as baryon junctions or hydro, while pion production is
dominated by the quenched pQCD spectra \cite{glvp}. A qualitatively different 
$v_2(p_T)$ behavior is therefore predicted for baryons and mesons.
Figure~\ref{fig:rich} compares $v_2$ of 
identified charged particles  
with pure hydrodynamical calculations \cite{hydro} 
for $p_T<3.5$ GeV/c.
In Figure~\ref{fig:lambda}, a similar comparison is presented for $K^0_s$ and
$\Lambda$. At $p_T<1$ GeV/c, the hydro calculations describe 
the mass dependence of the elliptic flow well. 
At $p_T>2$ GeV/c, 
$v_2$ of baryons may slightly exceed that of mesons \cite{glvp}. 

\section{CONCLUSIONS}

STAR has measured azimuthal anisotropies  as a function of
centrality in Au+Au collisions at \sqrtsNN=200 GeV of charged hadrons for $p_T<12$ GeV/c
and identified hadrons for $p_T<4$ GeV/c.
The experimental 
observation of large azimuthal anisotropies at high transverse momenta 
is the focus of ongoing theoretical investigation. Many different approaches
have been attempted to describe the data, such as combining hydrodynamical 
elliptic
flow with perturbative QCD including jet quenching \cite{wangglv}, 
parton cascade \cite{molnar}, minijets \cite{kovchegov}, and surface
emission \cite{shuryak}. A simultaneous quantitative description
of all the experimental data on high $p_T$ production at RHIC, including 
inclusive spectra suppression, saturation of $v_2$, and disappearance
of back-to-back jets in central collisions \cite{dave}, is needed. However, the
data clearly point in the direction of qualitatively new physics phenomena
in high $p_T$ particle production in nuclear collisions
at RHIC energies.

\end{document}